# A phenomenological thermodynamic potential for DyCo$_2$

Dongyan Zhang[1,*], Chao Zhou[2], Sen Yang[2], Zhimin Li[1], Xiaoping Song[2]

[1]School of Advanced Materials and Nanotechnology, Xidian University, Xi'an 710126, P.R. China

[2]MOE Key Laboratory for Nonequilibrium Synthesis and Modulation of Condensed Matter, Xi'an Jiaotong University, Xi'an 710049, P.R. China

**Abstract:** A phenomenological thermodynamic potential was constructed based on the symmetry analysis and property characterization of bulk DyCo$_2$. An eight-order polynomial of Landau expansion was employed to describe the thermodynamic behavior of DyCo$_2$. Several properties were reproduced including ferromagnetic transition temperature, magnetization curve, temperature dependence of magnetization. The transition behavior was analyzed via the thermodynamic potential. The Landau phenomenological thermodynamic model predicts the correct metamagnetic transition near the Curie temperature.

[*] Author to whom correspondence should be addressed
Email: zhangdongyan@xidian.edu.cn

1. Introduction

Ferromagnetic $DyCo_2$, one of the $RM_2$ (R=rare earth, M=transition metal) compounds with Laves phase[1,2], has been extensively studied for many decades due to its capability in magnetostriction[3,4] and magneto-refrigeration[5–7]. It exhibits large anisotropic magnetostriction and magnetic entropy change during phase transition[8–10]. Over the past decades, the good performance of magnetostriction or magnetocaloric in $DyCo_2$ has been explained via the exchange interaction between Dy, Co, and their neighbor atoms[11,12]. The large anisotropic magnetostriction is proposed to originate from the strong magnetoelastic coupling, which results in a remarkable shift of neighboring ion's positions when external magnetic field was applied[3]. The remarkable entropy change upon transition is attributed to the high magnetic moment for Co atoms which was induced by exchange interaction with Dy moments[5]. The micro-mechanism provides insight into how the behavior of atoms ordained the macroscopic performance. However, it is difficult to make a quantitative description of magnetostriction or magnetocaloric via the calculation of exchange interaction. Furthermore, no specific phenomenological thermodynamic theory was constructed to describe the magnetic properties and explain the transition behavior. This could be attributed to a long-term misunderstanding[4,13] that the ferromagnetic transition involves no structural change as a whole. This makes it difficult to expense the Landau polynomial under the specific symmetry, parallel to the Landau-Devonshire theory in ferroelectric system[14,15].

With the progress in ferromagnetic transition and high-resolution structure analysis, the noncubic symmetry ferromagnetic phase of $DyCo_2$ and a simultaneous structural change at ferromagnetic transition temperature in $DyCo_2$ was observed[4,13]. The simultaneous structural change in $DyCo_2$ demonstrated a paramagnetic-cubic to ferromagnetic-tetragonal phase transition at Curie temperature[13,16]. This provides us to develop a Landau polynomial with specific parameters for $DyCo_2$ based on the structure analysis and property characterization. In this work a thermodynamic description of the Landau theory was developed. It reproduces quite well most of the ferromagnetic properties including the phase transition temperature and metamagnetic transition behavior. The reproduced magnetic properties were compared to experimental values. The Landau potential predicts the correct metamagnetic transition near the Curie temperature and provides a thermodynamic analysis during this transition.

## 2. Phenomenological thermodynamic potential

In a phenomenological description of the ferromagnetic phase transition in $DyCo_2$, the spontaneous magnetization **M**= ($M_1$, $M_2$, $M_3$) is chosen as the order parameter. The Landau potential is expanded as a polynomial of the magnetization components $M_i$ ($i$=1, 2, 3)[17]. In this work, we employed an eighth-order polynomial for the Landau potential, it is

$$\begin{aligned}
G = & a_1(M_1^2 + M_2^2 + M_3^2) \\
& + b_{11}(M_1^4 + M_2^4 + M_3^4) + b_{12}(M_1^2 M_2^2 + M_2^2 M_3^2 + M_3^2 M_1^2) \\
& + c_{111}(M_1^6 + M_2^6 + M_3^6) + c_{112}[M_1^2(M_2^4 + M_3^4) + M_2^2(M_1^4 + M_3^4) + M_3^2(M_1^4 + M_2^4)] + c_{123} M_1^2 M_2^2 M_3^2 \\
& + d_{1111}(M_1^8 + M_2^8 + M_3^8) + d_{1112}[M_1^2(M_2^6 + M_3^6) + M_2^2(M_1^6 + M_3^6) + M_3^2(M_1^6 + M_2^6)] \\
& + d_{1122}(M_1^4 M_2^4 + M_2^4 M_3^4 + M_3^4 M_1^4) + d_{1123}(M_1^4 M_2^2 M_3^2 + M_1^2 M_2^4 M_3^2 + M_1^2 M_2^2 M_3^4) \\
& - H_1 \cdot M_1 - H_2 \cdot M_2 - H_3 \cdot M_3
\end{aligned} \quad (1)$$

where *a, b, c, d* are coefficients and **M** is the magnetization that is the order parameter, **H** is the external magnetic field, **G** is the Landau potential. *a* is linearly dependent on temperature and obeys the Curie-Weiss law.

The coefficients could be obtained by fitting to the ferromagnetic transition behavior from paramagnetic cubic phase to ferromagnetic tetragonal phase[13,16,18], and to the spontaneous magnetization of the tetragonal phase. From the experimental results, the ferromagnetic transition from a paramagnetic cubic phase to ferromagnetic tetragonal phase occurred at ~143K[4] (~145K[6]), and no other ferromagnetic transition accompanying with structural change was observed. Therefore, the magnetization conditions (in the absence of an external field) in the two stable phases of $DyCo_2$ are:

$$M_1^2 = M_2^2 = M_3^2 = 0 \quad \text{paramagnetic cubic m-3m} \quad (a)$$

$$M_1^2 = M_2^2 = 0, M_3^2 \neq 0 \quad \text{tetragonal 4/mmm} \quad (b)$$

The Landau polynomial could be simplified under tetragonal symmetry to:

$$G = a_1 M_3^2 + b_{11} M_3^4 + c_{111} M_3^6 + d_{1111} M_3^8 - H_3 \cdot M_3 \tag{2}$$

for the ferromagnetic phase, the first derivative of G,

$$\frac{\partial G}{\partial M_3} = H_3 \tag{3}$$

gives the filed component **$H_3$** (taken to be zero). These equations can then be solved to yield the spontaneous values of **$M_3$**. If the coefficients in Landau polynomial (Eq. 2) and their temperature dependencies are known, the function can be used to derive all the magnetic properties and their ranges of stability in each phase. The coefficients are given in Table I.

3. **Results and Discussion**

With the coefficients listed in Table I, the potential in Eq. (2) yields the transition temperature: $T_{Curie}$=133.87K for $DyCo_2$ single crystals under the stress-free condition. The Curie temperature in this work was compared to the experimental results in Table II. An obvious difference was observed in Table II between theoretical and experimental results. The theoretical Curie temperature for $DyCo_2$ in this work was obtained via solving the simultaneous Eq. (2) and (3) under the zero external field. However, the experimental Curie temperature was obtained via measured M-T curve and the value was determined by the intersection of extension line in M-T curves[4], as shown in Figure 1. The theoretical transition temperature, denoted as $T_{c\_theoretical}$, divided two thermodynamics stable states of $DyCo_2$, which is obtained by solving the thermodynamical equilibrium equations (Eq. (2) and (3)). The experimental transition temperature, denoted as $T_{c\_experimental}$, distinguished two magnetization region in M-T curve, which is obtained by magnetic characterization. The difference between $T_{c\_theoretical}$ and $T_{c\_experimental}$ is due to the metastable state above $T_{c\_theoretical}$. Figure 2 shows the free energy near transition temperature. When the temperature, T, is lower than $T_{c\_theoretical}$, the magnetization additional item reduces the free energy. Therefore, the system prefers the stable point as shown in Figure 2, energetically. As a result, ferromagnetic was manifested experimentally. When the temperature is $T_{c\_theoretical}$ <T< $T_{c\_experimental}$, theoretically the system should be paramagnetic and the magnetization is zero. However, a metastable point where magnetization is not zero exists in this temperature region, so a nonzero magnetization could be measured experimentally. As T> $T_{c\_experimental}$, the metastable point disappears, the system is paramagnetic in both theoretical analysis and experimental measurement. Therefore, $T_{c\_theoretical}$ is lower than $T_{c\_experimental}$, and the magnetization is nonzero when the temperature is little higher than the theoretical Curie temperature.

The predictions from the model function for magnetic susceptibility, $\chi_{33}$, of $DyCo_2$ are given in Figure 3. In the Landau theory it is assumed that below the Curie point, the magnetic susceptibility could be calculated according to its definition[19]

$$\chi = \frac{\partial M}{\partial H} \tag{4}$$

above the Curie point the magnetic susceptibility is

$$\frac{1}{\chi} = a_0(T - T_0), a_0 = 7.3 \times 10^{-4}, T_0 = 131.19 \tag{5}$$

which captures the Curie-Weiss behavior as shown in Figure 3. This provides additional support for our linear temperature ansatz for $a_1$.

To check whether the phenomenological model is applicable for $DyCo_2$ magnetic properties, the magnetization curves at 138K were calculated to be compared with measured value in Figure 4a, where the calculated curve is consistent with the measured value. Both display a behavior with typical soft ferromagnets. An inflection point (denoted as critical field in Figure 4a) was observed in both calculated and measured curves, which indicates a metamagnetic behavior. The magnetization jump occurs with increasing field across the critical point. The presence of a metamagnetic transition in $DyCo_2$ is also studied in the free energy drawing, as shown in Figure 4b. When the external field is weak, the magnetization is near zero. Therefore, the system manifested paramagnetic (black and red dash curve in Figure 4b). As the external field increase across the critical point, the magnetization is much bigger than that in the paramagnetic state. Therefore, the system manifested ferromagnetic (blue curve in Figure 4b).

The presence of a metamagnetic transition in $DyCo_2$ is seen not only in the magnetization isotherms but also in the Arrott plots [20]($M^2$ vs H/M) as shown in Figure 5, which are usually used to determine the transition order[21] for metamagnetic behavior with increasing field. If an inflection point is found in Arrott plot curve, it indicates a first-order transition. Otherwise, it indicates a second-order transition. As an example, the Arrott plots for $DyCo_2$ at 130K, 134K, 138K, 142K, 146K, and 150K are shown in Figure 5. When the temperature is not above Curie temperature (130K, 134K), $DyCo_2$ manifests ferromagnetic state. With the external magnetic field increasing, no metamagnetic transition was observed. When the temperature is above Curie temperature, $DyCo_2$ undergoes a metamagnetic transition with external magnetic field increasing. The existence of inflection point in Arrott plots ($M^2$ vs H/M red dash curves at 146K, and 150K in Figure 5) determines the order of metamagnetic transition. The absence of inflection point in Arrott plot ($M^2$ vs H/M red dash curves at 138K and 142K in Figure 5) suggests the occurrence of a second-order metamagnetic transition from paramagnetic to ferromagnetic. The metamagnetic behavior plays a critical role in determining the order of magnetic transition which influences directly the thermodynamic properties.

According to Shimizu's theory[22], the transition of $DyCo_2$ can be explained by an s-d model, where localized spins of rare earth atoms interact with the itinerant d electrons of Co atoms. Thus, the order parameter, magnetization, in Landau polynomial related to the integral of localized spins. Furthermore, the coefficients of Landau polynomial in Eq. (2) are modified by the exchange interaction. The nature of the transition is determined by the dependence of Landau energy on the magnetization and temperature. The equilibrium condition as shown in Eq. (3) allows one to relate the magnetization and field through the state equation:

$$\frac{H}{M} = 2a_1 + 4b_{11}M_3^2 + 6c_{111}M_3^4 + 8d_{1111}M_3^6 \quad (6)$$

The coefficient $a_1$ and $b_{11}$ depend on the temperature with respect to thermal variations of spin fluctuation amplitude and can be determined by fitting the isothermal magnetization data using Arrott plot and Eq. (5). The sign of coefficients will determine the nature of the paramagnetic-ferromagnetic transition. $a_0$ and $c_{111}$ are both positive in this phenomenological theory for $DyCo_2$. The magnetic transition order is governed by the sign of $b_{11}$. If it is negative, a 1$^{st}$ order transition takes place and its result the magnetization jumps discontinuously. Otherwise, a

$2^{nd}$ order transition occurs, and the magnetization change continuously. Therefore, in the absence of an external field, the magnetization jumps discontinuously at $T_c$ (as shown in Figure 1 and 2 ), and denoted as $1^{st}$ order paramagnetic-ferromagnetic transition. In the case of external field, the magnetization change continuously near critical field (as shown in Figure 4), and this is $2^{nd}$ order metamagnetic transition; the magnetization change discontinuously near critical field, which indicates an existence of inflection point in Arrott plot, and this is $1^{st}$ order metamagnetic transition. By extracting the critical point in the magnetization curves at different temperature, a phase diagram of $DyCo_2$ under external magnetic field was drawn, as shown in Figure 6.

## 4. Conclusion

A phenomenological thermodynamic potential was constructed based on the symmetry analysis and properties characterization of bulk $DyCo_2$. An eight-order polynomial of Landau expansion was employed to analyze the thermodynamic behavior of $DyCo_2$ and to reproduce its properties. The reproduced properties are consistent with the experimental value. Furthermore, the Landau phenomenological thermodynamic model in this work predicts the correct metamagnetic transition near the Curie temperature and deduces the phase diagram of $DyCo_2$ under external filed.

**Acknowledgments**

This research was supported by the Fundamental Research Funds for the Central Universities (Grant No. XJS16028), and National Science Foundation of China (Grant No. 51601140, 51471125, 51371134).

**Tables**

Table I. Coefficients of Landau potential in Eq. (1) where T is temperature in K

| $a_1$ | $b_{11}$ | $c_{111}$ | $d_{1111}$ |
|---|---|---|---|
| 0.00073*[T-131.19] | 0.4*10$^{-7}$*[T-148.38] | 0.78*10$^{-11}$ | 0.62*10$^{-14}$ |

Table II. Comparison of Curie temperature between theoretical and experimental results

| transition | this work | experimental results |
|---|---|---|
| cubic to tetragonal (Curie temperature) | 133.87K | 143K[5, 7], 145K[5, 7] |

**Captions**

**Figure 1.** Comparison of normalized Magnetization between experimental data[5, 7] and theoretical value

**Figure 2.** Free energy near transition temperature.

**Figure 3.** Magnetic permeability of $DyCo_2$ in the corresponding ferromagnetic tetragonal phase, where $\mu_{33}$ is along the magnetization direction.

**Figure 4.** (a).Theoretical H-M curve and experimental H-M loop at 138K. (b) Free energy dependent magnetization at 138K with different external field.

**Figure 5.** Arrott plots for $DyCo_2$ .

**Figure 6.** Phase diagram under external magnetic field.

**Figures**

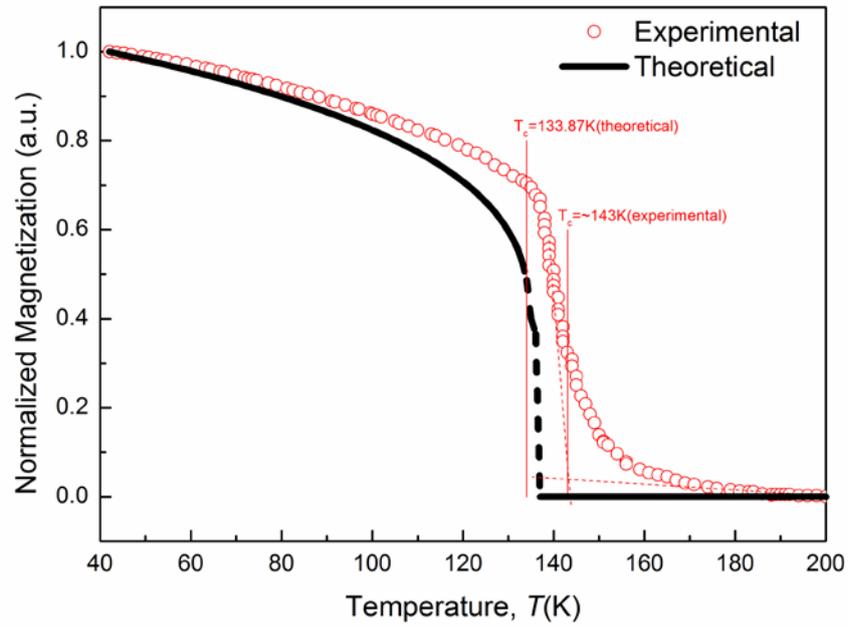

Figure 1. Comparison of normalized Magnetization between experimental data[5, 7] and theoretical value

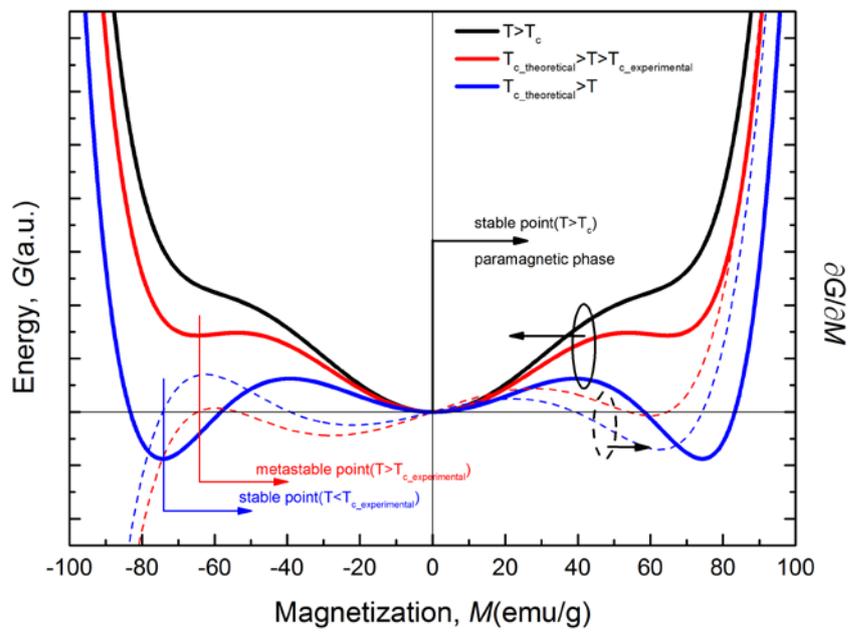

Figure 2. Free energy near transition temperature.

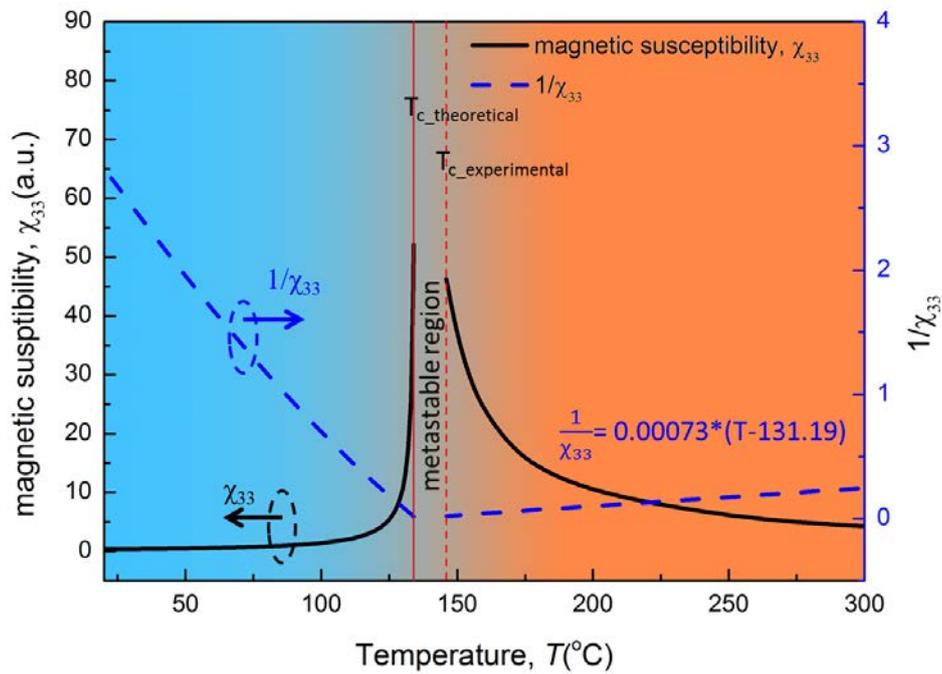

Figure 3. Magnetic susceptibility of DyCo$_2$ in the corresponding ferromagnetic tetragonal phase, where $\mu_{33}$ is along the magnetization direction.

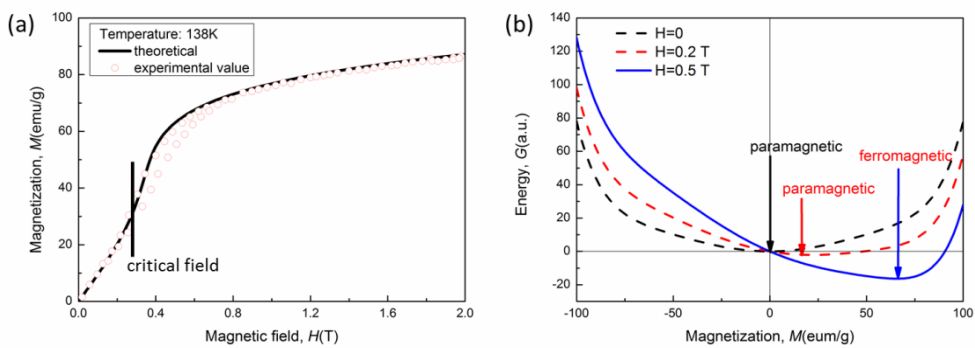

Figure 4. (a). Theoretical H-M curve and experimental H-M loop at 138K. (b) Free energy dependent magnetization at 138K with different external field.

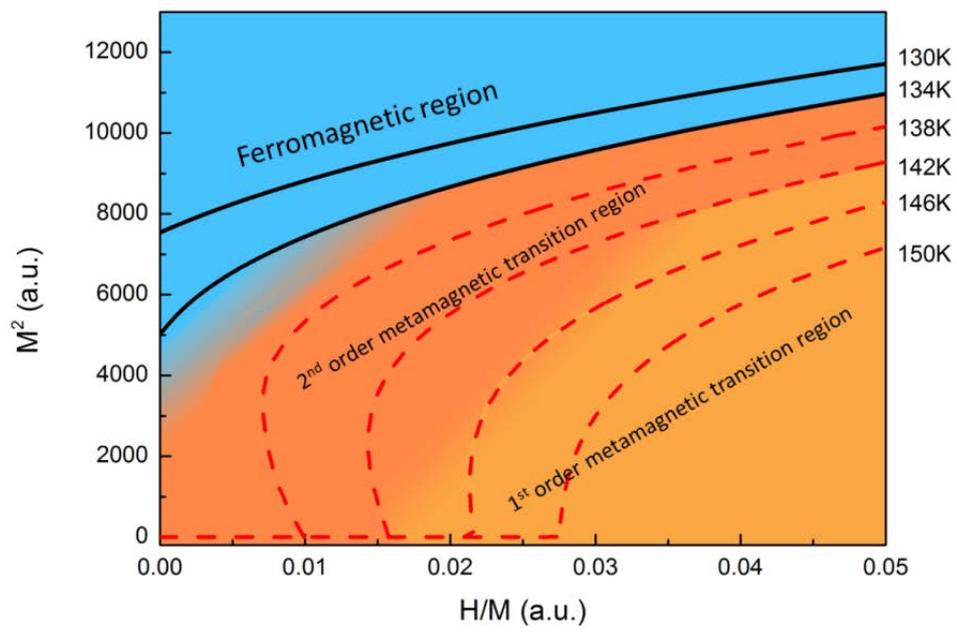

Figure 5. Arrott plots for $DyCo_2$.

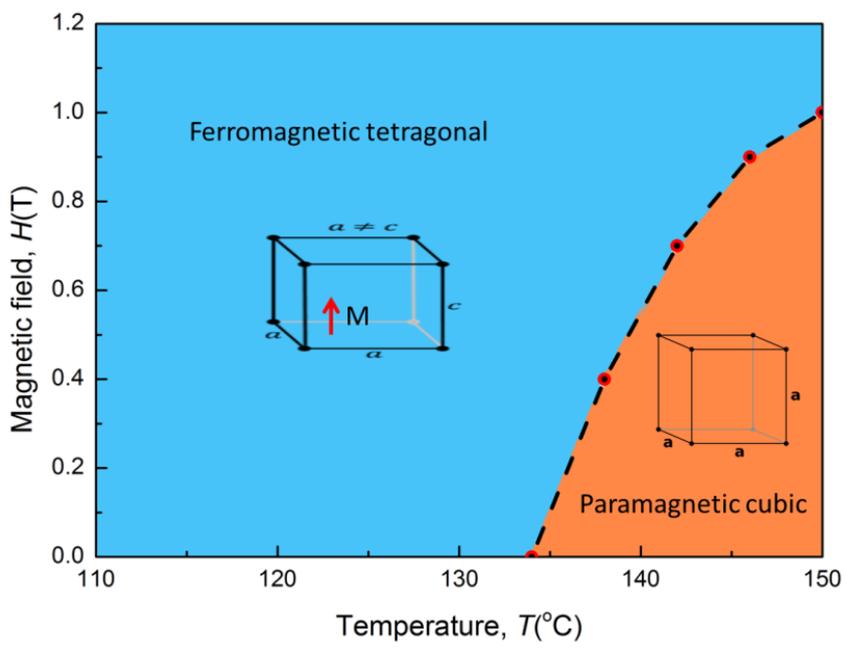

Figure 6. Phase diagram under external magnetic field.